# Hessian sparsity-constrained self-supervised network for near-infrared single-photon single-pixel imaging

Yao Wang[1], Muchen Zhu[1], Linjun Zhai[1], Huiyuan Zhang[1], Junnan Chen[1],
Yiming Yu[1], Zhaohua Yang[2,3], Baolei Liu[2,3*], and Fan Wang[1*]

[1]School of Physics, Beihang University, Beijing, 100191, China
[2]School of Instrumentation and Optoelectronics Engineering, Beihang University, Beijing, 100191, China
[3] Hangzhou International Innovation Institute, Beihang University, Hangzhou, 311115, China
*Correspondence: liubaolei@buaa.edu.cn; fanwang@buaa.edu.cn

**Keywords:** single-pixel imaging, single-photon imaging, near-infrared imaging, self-supervised deep learning

**Abstract:** Near-infrared (NIR) imaging has emerged as an important technology for night vision, remote sensing, and biological imaging, yet conventional array-detector-based systems are often limited by insufficient sensitivity, high cost, and substantial dark noise. Single-pixel imaging (SPI) offers an attractive alternative, enabling single-photon-level NIR imaging by using a cost-effective single-element detector. Nevertheless, SPI remains restricted by photon noise, leading to degraded imaging quality and limited frame rate under extremely low photon flux conditions. Here, we present a Hessian sparsity-constrained self-supervised network ($HS^3N$) for single-photon NIR SPI, which can suppress noise and enable high-fidelity and real-time imaging under ultra-low illumination conditions. The $HS^3N$ integrates the physical forward model of SPI with an untrained neural network regularized by both sparsity priors and Hessian-based structural constraints, enabling effective noise suppression while preserving structural fidelity and continuity. Both simulated and experimental results demonstrate that $HS^3N$ enables high-fidelity reconstructions under ultra-low NIR photon levels down to ~0.01 photons per pixel. Furthermore, we demonstrate its dynamic capability by monitoring the dynamic evolution and detachment of infrared-absorbing droplets, at a frame rate of ~20 Hz under ~0.19 photons per pixel, highlighting its potential for high-sensitivity infrared inspection. The proposed reconstruction framework paves the way for practical NIR imaging in extreme low-light conditions, which can be extended to visible, mid-infrared or terahertz imaging, offering broad potential for photon-efficient sensing across a wide spectral range.

## 1. Introduction

Near-infrared (NIR) imaging plays an important role across a wide range of applications, including remote sensing, night vision, and biological imaging, owing to its ability to capture thermal radiation and reveal information beyond the visible spectrum [1][2][3][4]. In recent years, single-photon NIR imaging has attracted growing attention for its ability to operate under extremely low photon flux, enabling high-sensitivity detection in photon-starved environments. The unique advantages of photon-level sensitivity and high timing resolution make it well suited for a wide range of applications, including three-dimensional light detection and ranging, quantum imaging and bio-photonics [5][6][7]. However, NIR imaging systems often rely on focal plane array detectors, which are often expensive and involve complex fabrication processes, particularly in low-photon-flux regimes.

Single-pixel imaging (SPI) has emerged as a cost-effective solution to conventional imaging by replacing the array detector with a single-element detector, followed by the computational post-reconstruction [8]. In SPI, a sequence of structured patterns generated by the spatial light modulator is utilized to illuminate objects, and the single-pixel detector (SPD) records corresponding reflected/transmitted light intensities [9]. The object is then reconstructed from the correlations between the illumination patterns and the measured signals using algorithms such as linear correlation [10], and compressive sensing [11][12][13], and deep learning [14][15][16]. Benefiting from its broadband response, low cost, and simple yet robust architecture, SPI has been extended across a wide spectral range [17][18][19][20]. SPI has also been applied to single-photon imaging under extremely low-light conditions by combining with photon-counting detectors [21][22][23], which significantly reduces hardware complexity and cost, especially in invisible spectral band [24][25][26]. NIR single-photon SPI has showcased great potential for broad applications in long-distance sensing, deep-tissue imaging, and night vision under ultra-low infrared irradiation, owing to its simplicity, low-cost system and robustness [27][28]. However, NIR single-photon SPI remains fundamentally limited by low SNR under ultra-low illumination, where severe photon noise degrades both reconstruction quality and imaging speed, thereby hindering high-sensitivity, real-time NIR imaging in photon-starved scenarios.

In this paper, we propose a Hessian sparsity-constrained self-supervised network ($HS^3N$) for near-infrared single-pixel imaging, toward robust and high-quality image reconstruction in photon-starved conditions. The $HS^3N$ introduces a composite loss function, including a data fidelity term $\mathcal{L}_{\mathrm{Data}}$ and a joint regularization term $\mathcal{L}_{\mathcal{R}}$ where the joint regularization term combines $\ell_1$-norm sparsity and Hessian norm to promote sparsity and continuity. This dual-constraint strategy allows the model to effectively suppress noise while preserving structural details. We then validate $HS^3N$ through both simulations and experiments, which showcase its strong performance under ultra-low photon flux levels. The experimental results demonstrate that $HS^3N$ achieves superior reconstruction quality at photon levels as low as ~0.01 photons per pixel under near-infrared illumination of 852 nm. Moreover, dynamic imaging of infrared-absorbing droplet formation is demonstrated at a frame rate of ~20 Hz under ~0.19 photons per pixel. These results highlight the effectiveness of $HS^3N$ for high-sensitivity imaging under photon-starved conditions and its potential for real-time NIR imaging applications.

## 2. Results

The schematic of NIR single-photon SPI system is illustrated in Fig. 1a, which mainly consists of a NIR laser source, a digital micromirror device (DMD), and a photon-counting detector. The conceptual architecture of the proposed Hessian sparsity-constrained self-supervised network ($HS^3N$) is illustrated in Fig. 1b. $HS^3N$ is a self-supervised framework without pre-training that treats the reconstruction as an iterative optimization process guided by the physical model of SPI, as shown in Fig. 1c. The neural network $\mathcal{N}_\theta$ (based on an encoder-decoder U-Net architecture) takes an initial differential ghost imaging (DGI) $x_{\text{in}}$ (or a random input) and maps it to a predicted image $x_{\text{out}}$. According to the output image, the predicted 1-D signals $I_{\text{pred}}$ can be generated by performing Hadamard product between the illumination Hadamard patterns $H = [\mathrm{h}_1, \mathrm{h}_2, \dots, \mathrm{h}_\mathrm{N}]$ and $x_{\text{out}}$ through a forward physical model $\mathcal{F}$:

$$\begin{gathered} I_{\text{pred}} = \mathcal{F}(H, x_{\text{out}}), \\ \left[I_{\text{pred}}\right]_i = \sum (h_i \odot x_{\text{out}}), \end{gathered} \tag{1}$$

where $i$ is the index of the illumination patterns and measurements and $\odot$ represents Hadamard product. Then the first data fidelity loss is calculated as the mean squared error (MSE) between the predicted signals $I_{\text{pred}}$ and the experimentally measured signals $I_{\text{real}}$:

$$\mathcal{L}_{\text{Data}} = \left\| I_{\text{pred}} - I_{\text{real}} \right\|_2^2. \tag{2}$$

Here, we introduce a joint regularization loss term $\mathcal{L}_{\mathcal{R}}$ that combines low-order and high-order image prior to suppress Poisson noise inherent in photon-starved regime:

$$\mathcal{L}_{\mathcal{R}} = \lambda_1 \| x_{\text{out}} \|_1 + \lambda_2 \| \nabla^2 x_{\text{out}} \|_F^2, \tag{3}$$

where $\lambda_1$ and $\lambda_2$ are weight coefficients. Specifically, the $\ell_1$-norm sparsity term ($\| x_{\text{out}} \|_1$) promotes sparsity in the reconstructed image, effectively suppressing noise-induced fluctuations while preserving dominant structural components [29]. The Hessian Frobenius norm ($\| \nabla^2 x_{\text{out}} \|_F^2$) acts as a high-order smoothness prior, promising smooth and continuous structures while suppressing unwanted fluctuations [30]. The network parameters $\theta^*$ are optimized by minimizing the total loss function $\mathcal{L}$, which is formulated as:

$$\theta^* = arg\,\min_{\theta} (\mathcal{L}_{Data} + \mathcal{L}_{\mathcal{R}}). \tag{4}$$

During the implementation, a U-Net architecture is employed. The input is initialized using the DGI image. The reconstructed image exhibits improved quality with reduced noise as $\ell_1$-norm sparsity decreases during the optimization, as shown in Fig. 1d. The optimization is carried out using the Adam optimizer, with an initial learning rate of $10^{-2}$ and an exponential decay schedule. The batch size is set to 1, and the reconstruction is performed over 100 iterations. The weighting parameters for the $\ell_1$-norm and Hessian-based regularization are set to $10^{-5}$. All experiments are conducted on a standard laptop with a Core i7-12800HX processor and an Nvidia RTX 4070 GPU.

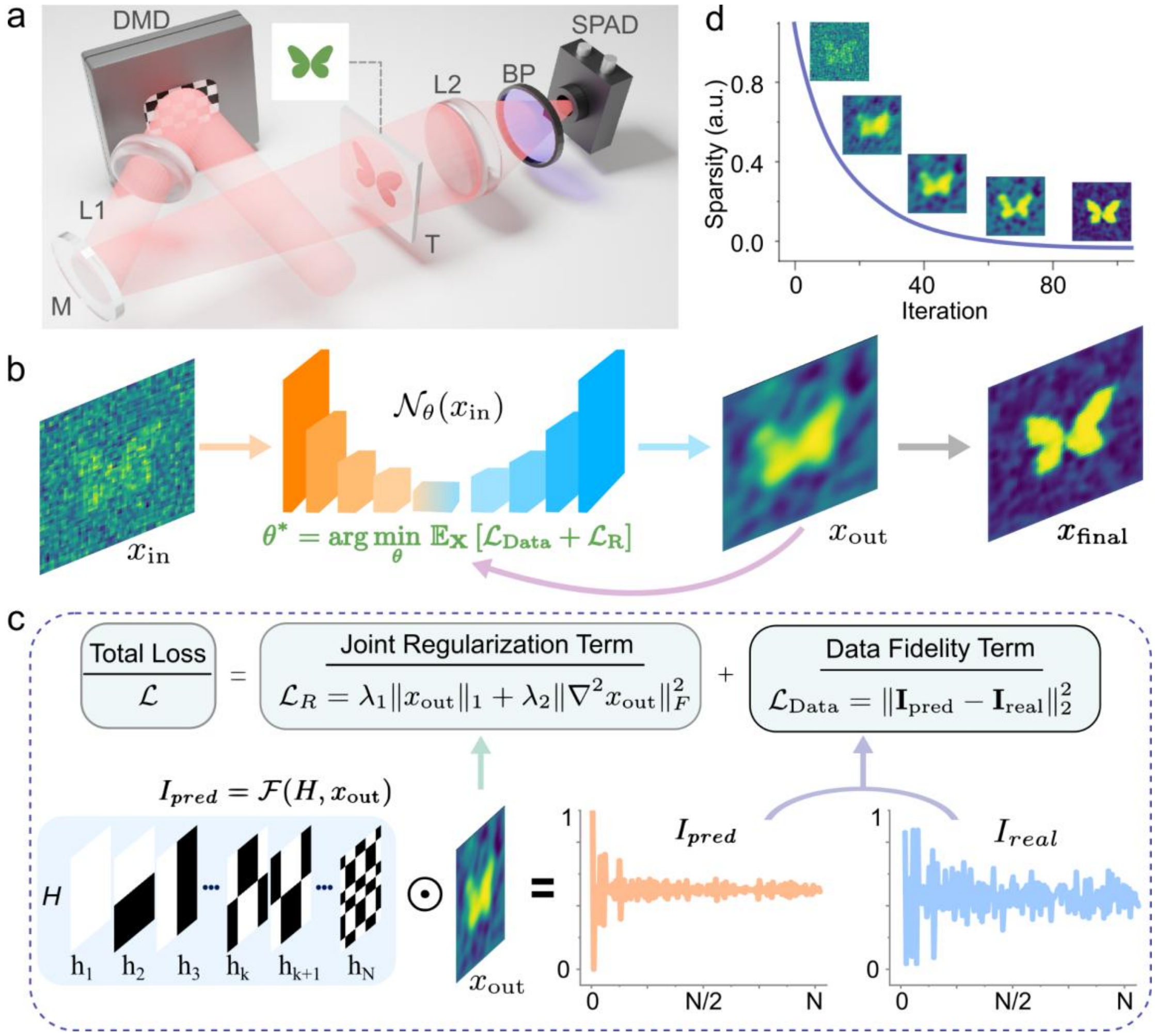

**Figure 1**. Principles of the proposed Hessian sparsity-constrained self-supervised network ($HS^3N$) for single-photon single-pixel imaging. **(a)** Experimental setup of SPI system. L1–L2, lenses; BP, band-pass filter; DMD, digital micromirror device; M, plane mirror; T, target; SPAD, single-photon avalanche diode. **(b)** Iterative optimization workflow of $HS^3N$. The neural network $\mathcal{N}_\theta$ transforms the initial input $x_{in}$ into the reconstructed image $x_{out}$. The network parameters $\theta$ are optimized by minimizing the total loss function. **(c)** Loss formulation and physical model. The total loss $\mathcal{L}$ integrates a data fidelity term $\mathcal{L}_{\text{Data}}$ and a joint regularization term $\mathcal{L}_{\mathcal{R}}$ (consisting of $\ell_1$-norm sparsity and Hessian norm $\|\nabla^2 x_{\text{out}}\|_F^2$ constraints). The predicted signal $I_{\text{Pred}}$ is generated via the forward physics model $\mathcal{F}$ by performing Hadamard product between the illumination patterns $[h_1\, h_2\, ...\, h_N]$ and $x_{out}$, then the MSE between $I_{\text{pred}}$ and the experimental measurement $I_{\text{real}}$ is as $\mathcal{L}_{\text{Data}}$. **(d)** Sparsity ($\|x_{\text{out}}\|_1$) of the reconstructed image as a function of iteration.

To evaluate the performance of the proposed $HS^3N$, we first conduct numerical simulations under photon-starved conditions. Five representative targets with a size of 64 × 64, including text ("OPT", "25"), geometric shapes, a butterfly, and microtube, are utilized to test the proposed method's robustness, as illustrated in left panel of Fig. 2. Poisson noise was numerically introduced into the measurements under various average photon levels of ~0.0049, 0.0065, and 0.022 photons per pixel each detection. The results show that $HS^3N$ achieves better reconstruction quality with less noise across all photon levels compared to traditional DGI and ghost imaging (GI) using deep neural network constraint (GIDC) under a fixed sampling ratio of 40% [15]. At the ultra-low illumination level of 0.0049 average photons per pixel, DGI results are dominated by dense noise, making the targets almost unrecognizable, while GIDC image is buried in noise and fails to resolve the targets. In contrast, HS$^3$N effectively suppresses the background noise while maintaining primary structures, thereby resolving the targets clearly. As increasing the photon intensity, all three methods improve the quality of reconstruction images. But HS$^3$N consistently delivers superior results with clearer structures and reduced noise, compared to the other two methods. Furthermore, the average structural similarity (SSIM) of HS$^3$N is significantly higher than that of the baseline methods across all photon levels, as shown at the bottom of Fig. 2 and Table 1. These results showcase the effectiveness of the joint $\ell_1$-norm and Hessian regularization strategy in combating severe Poisson noise.

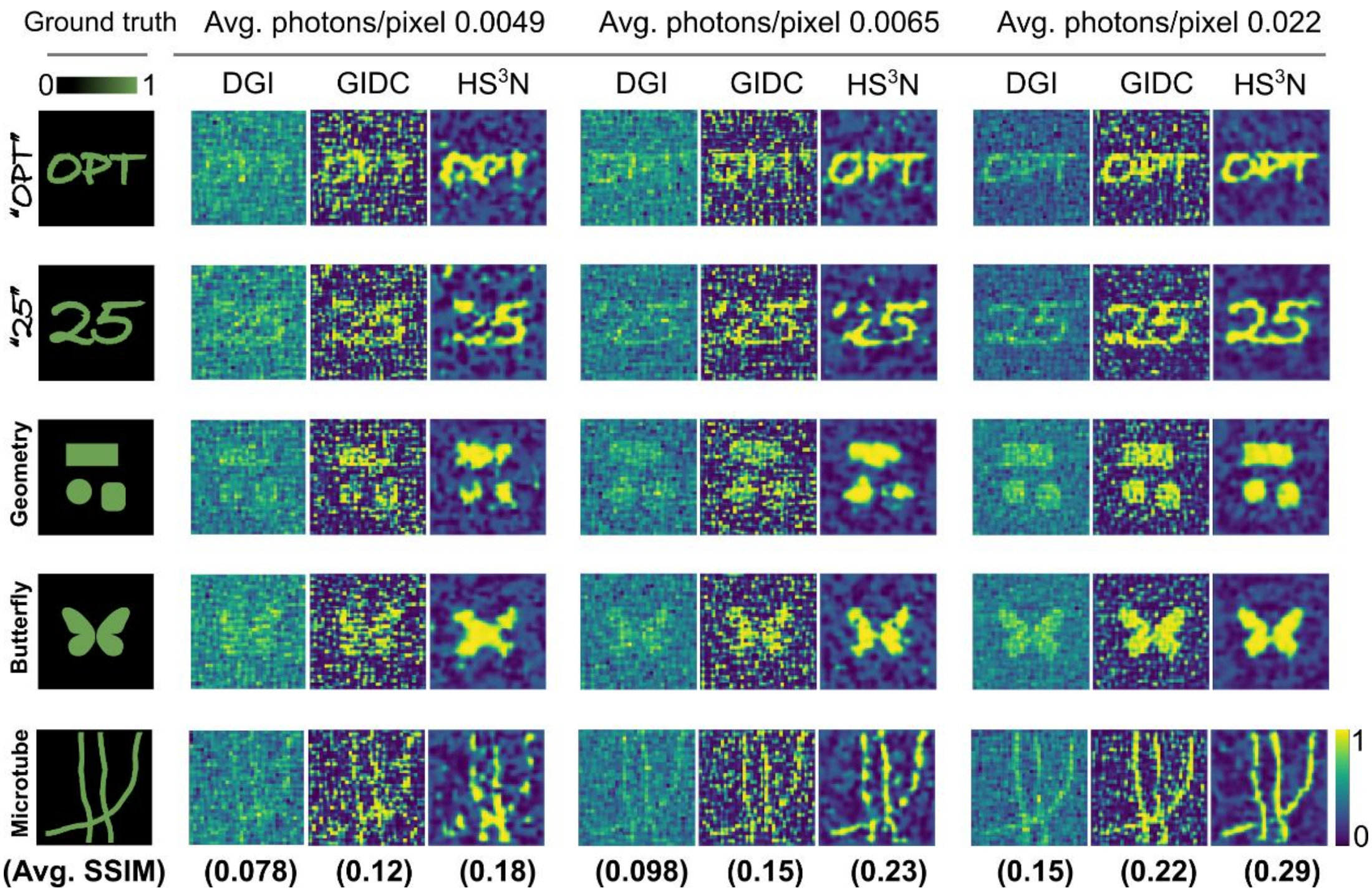

**Figure 2**. Simulated comparison of DGI, GIDC and HS$^3$N under varying light levels. Poisson noise was applied to simulate different photon-count conditions. All reconstructions used a sampling ratio of 40% across five test targets: "OPT", "25", Geometry, Butterfly, and Microtube.

**Table 1.** SSIM comparison under different methods and photon levels

| Photon levels | 0.0049 photons/pixel | | | 0.0065 photons/pixel | | | 0.022 photons/pixel | | |
|---|---|---|---|---|---|---|---|---|---|
| Targets | DGI | GIDC | Ours | DGI | GIDC | Ours | DGI | GIDC | Ours |
| "OPT" | 0.062 | 0.11 | **0.14** | 0.089 | 0.14 | **0.17** | 0.14 | 0.19 | **0.22** |
| "25" | 0.088 | 0.13 | **0.19** | 0.11 | 0.17 | **0.23** | 0.16 | 0.24 | **0.30** |
| Geometry | 0.083 | 0.11 | **0.18** | 0.079 | 0.098 | **0.21** | 0.13 | 0.20 | **0.28** |
| Butterfly | 0.063 | 0.088 | **0.16** | 0.062 | 0.087 | **0.17** | 0.11 | 0.17 | **0.24** |
| Microtube | 0.094 | 0.15 | **0.18** | 0.14 | 0.24 | **0.33** | 0.20 | 0.29 | **0.36** |

We then perform NIR single-photon SPI experiments with a standard USAF-1951 resolution target, as shown in Fig. 3a. The illumination source is a continuous-wave laser of 852 nm (MDL-III-852L, CNI). The laser beam is expanded to ~10 mm by a pair of lenses. The digital micromirror device (DMD, GmbH V-7001 ViALUX) is used to modulate the infrared light with the only central 32/64 rows of DMD. Then patterns are projected and magnified using a single projection lens (L1, $f$ = 50 mm). The NIR photons pass through the targets and are collected by a convex lens (L2, $f$ = 100 mm), filtered by a band-pass filter (MBF10-850-50, LBTEK). Then photons are guided by an optical fiber with a 200 μm core diameter (FCM2-PC-200L, JCOPTIX) and then detected by the single-photon avalanche diode (SPAD, SPCM-AQRH-14, Excelitas). The photon-count signals from SPAD are recorded by a data acquisition card (USB-6353, National Instruments), which is synchronized with the DMD trigger signal to ensure precise temporal alignment between the patterns and the measurements by using a purpose-built Python program. By varying the illumination time from 0.05 to 0.8 ms, we characterized the system under a gradient of ultra-low light intensities, ranging from 0.011 to 0.19 average photons per pixel. The side-by-side 64 × 64 reconstruction results of DGI, GIDC, and HS$^3$N across different photon levels with a sampling ratio of 40% are shown in Fig. 3b. All approaches show a consistent trend of improved reconstruction quality as increasing the photon level. But the reconstructions of DGI and GIDC are heavily degraded by background and shot noise, particularly under ultra-low photon conditions. Compared to the other two methods, HS$^3$N consistently outperforms the other two methods, producing clearer structures and significantly reduced noise, which demonstrates that HS$^3$N can effectively suppress Poisson noise while retaining the essential structural components of the target. Line profiles across the resolution elements (indicated by the white arrows in Fig. 3b) are compared in Fig. 3c, where HS$^3$N features higher contrast. These results demonstrate robustness and better reconstruction with reduced noise of HS$^3$N for high-fidelity NIR imaging under photon-starved conditions.

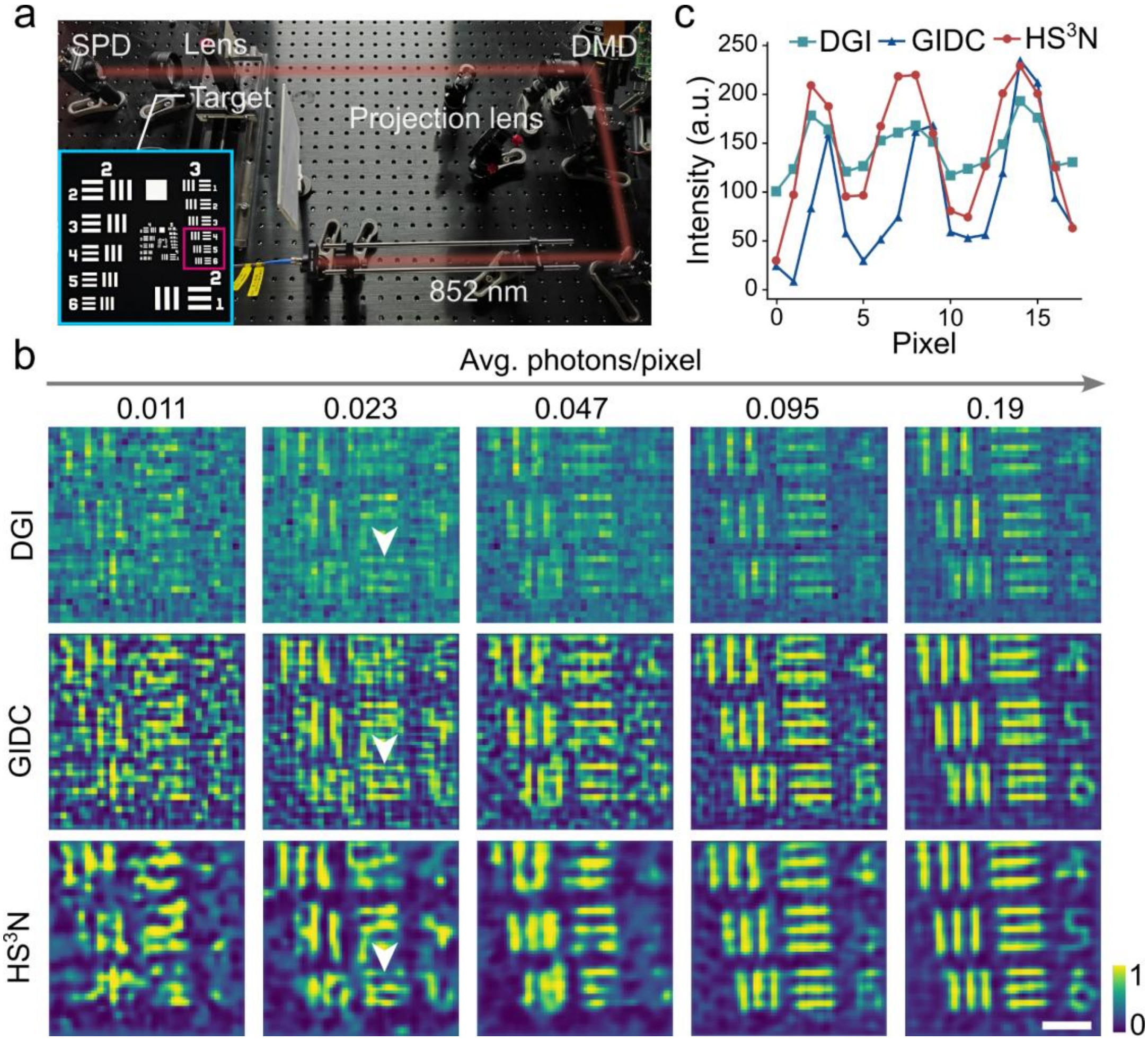

**Figure 3**. Experimental comparison of DGI, GIDC and HS$^3$N under photon-starved conditions with a standard target. **(a)** Photograph of the experimental system. A USAF-1951 resolution chart is used as the target. The inset shows the imaging region highlighted by the red box. **(b)** Comparison of reconstructed images from DGI, GIDC, and HS$^3$N across a range of average photon counts (0.011-0.19 photons/pixel), where GIDC and HS$^3$N were run for 100 iterations each. **(c)** Intensity profiles along the white arrows shown in (b). Scale bar, 5 mm.

To further evaluate the practicality of HS$^3$N, we conduct experiments on static targets under ultra-low infrared irradiation. Figure 4a illustrates the target configuration consisting of a randomly distributed infrared-absorbing ink spot deposited on a coverslip with an absorption peak at ~850 nm. The ink spot is transparent in the visible spectrum but exhibits high contrast under NIR illumination. The cyan circle indicates the region within the field of view (FOV). The infrared-absorbing spot, and the FOV are highlighted in purple and pink in the lower-right inset, respectively. Figure 4b presents a comparison of 64 × 64 reconstructions obtained by DGI, GIDC, and HS$^3$N under an average illumination level of ~0.27 photons per pixel and a sampling ratio of 20%, with 100 iterations used for GIDC and HS$^3$N. It can be seen that both DGI and GIDC are severely degraded by severe noise, leading to blurred structures and poor contrast. In contrast, HS$^3$N effectively suppresses photon noise and recovers the infrared-absorbing spot with significantly improved clarity. Moreover, we conducted experiments by using a needle tip without and with an infrared-absorbing droplet under average illumination levels of ~2.06 and 0.78 photons per pixel, respectively, as shown in Figs. 4c and 4e. The corresponding 32×32 results under a sampling ratio of 40% are illustrated in Figs. 4d and 4f. The needle profile and droplet region are clearly resolved by HS$^3$N, demonstrating the effectiveness of the proposed joint regularization strategy in preserving structural information while suppressing noise. By contrast, the other two methods yield reconstructions heavily degraded by noise. These results confirm that HS$^3$N enables robust and high-fidelity reconstruction for static NIR scenes, even under extremely low photon flux conditions.

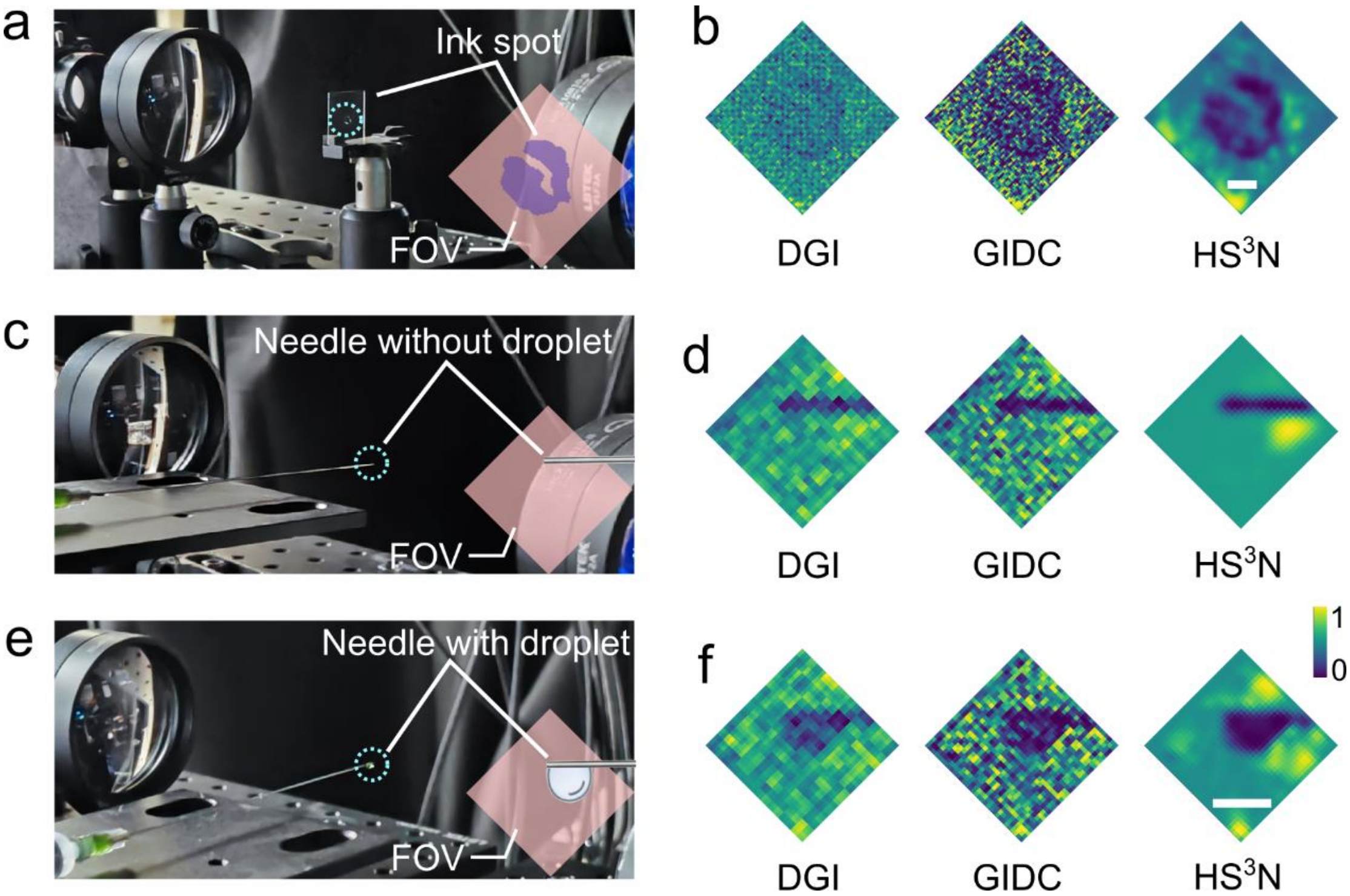

**Figure 4.** Experimental results with infrared-absorbing targets. **(a)** Photograph of the experimental target, a coverslip with a randomly distributed infrared-absorbing spot (cyan circle). **(b)** Reconstructed images of the target in (a) using three different reconstruction algorithms, DGI, GIDC and HS$^3$N. **(c)** Photograph of a syringe needle tip without an infrared-absorbing droplet. **(d)** Reconstructed images of the needle tip corresponding to (c), showing the needle morphology. **(e)** Photograph of a syringe needle tip with an infrared-absorbing droplet attached. **(f)** Reconstructed images of the needle tip corresponding to (e), in which the droplet is resolved as an additional feature adjacent to the needle body. Scale bar, 5 mm.

To validate the capability of HS$^3$N in dynamic scenarios, we perform real-time inspection of infrared-absorbing droplet formation and dripping at the needle tip. The photograph of the dynamic target, highlighted by the cyan circle, is shown in Fig. 5a. The droplet is generated by depressing the syringe plunger, and its evolution and detachment are dynamically imaged to capture the transient process. Figure 5b is the schematic of the evolution process of the droplet from initial formation to eventual detachment. The corresponding reconstructed image sequence is presented in Fig. 5c.

Despite operating at an average photon level of ~0.19 photons per pixel, $HS^3N$ successfully captures the temporal evolution of the droplet at a sampling ratio of 20% and a spatial resolution of 32 × 32 pixels. The growth of the droplet beneath the needle and the final detachment are clearly resolved from 0 ms to 350 ms with a frame rate of ~20 Hz, as highlighted by the dashed outlines in Fig. 5c. Initially, only the needle is visible, while the droplet detaches at 350 ms, as highlighted by the yellow arrow. Figure 5d shows the corresponding photon-count measurements over time, where fluctuations induced by the droplet evolution can be observed in the inset. Furthermore, quantitative analysis of the droplet area (Fig. 5e) shows a clear increasing trend over time, which is consistent with the physical process of droplet accumulation, demonstrating the reliability of the reconstruction. The experiment demonstrates that $HS^3N$ not only achieves high-quality reconstruction under ultra-low photon flux, but also enables reliable dynamic imaging, highlighting its potential for real-time NIR imaging applications.

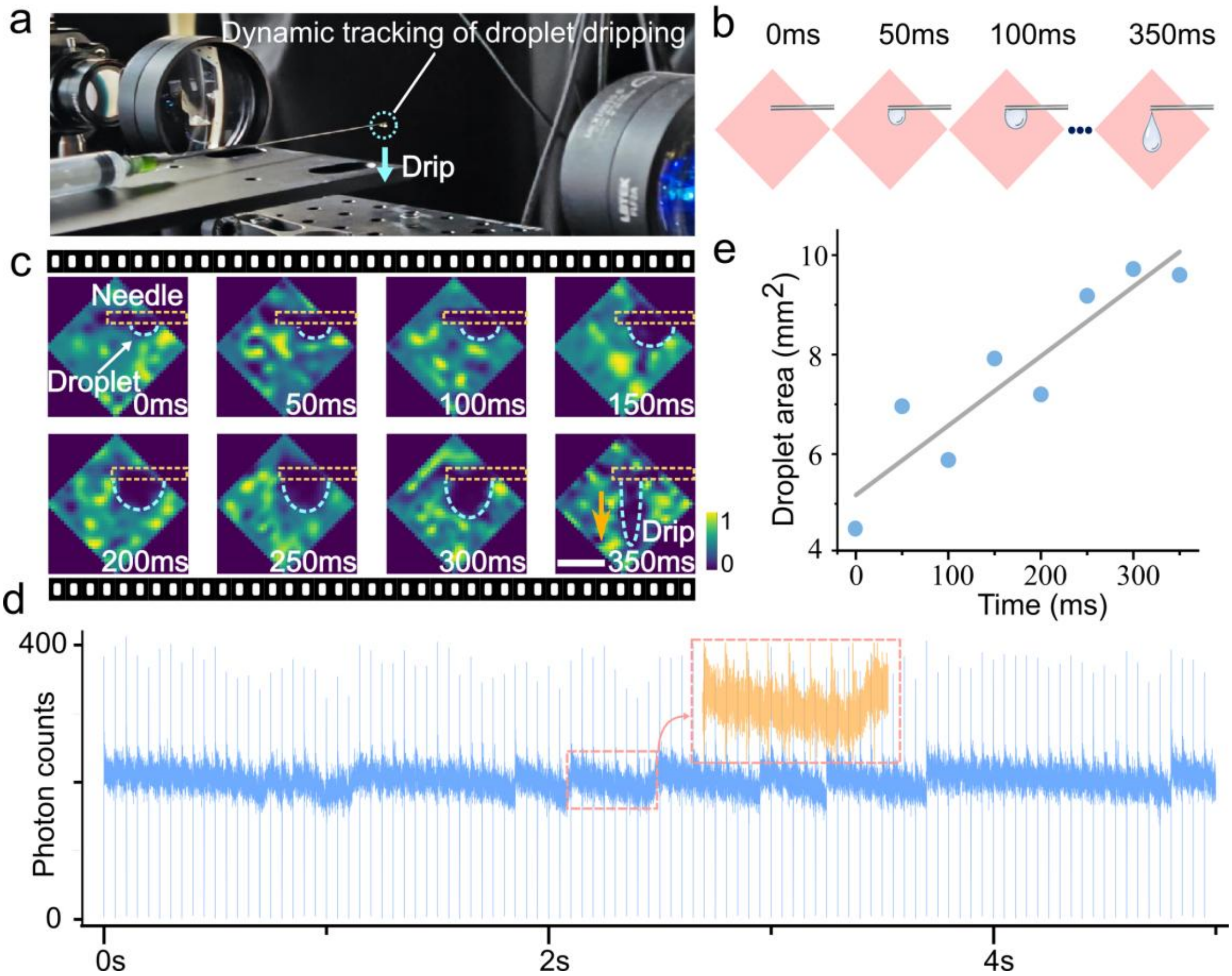

**Figure 5.** Dynamic tracking of infrared-absorbing droplet formation and dripping with $HS^3N$. **(a)** Photograph of the tracking droplet attached to the needle (cyan circle). We depress the syringe plunger to generate the droplets and let it drip to simulate the dynamic process. **(b)** Schematic illustration of the droplet evolution, from initial formation to detachment at the needle tip. **(c)** Time-lapse reconstructions of droplet formation and detachment, acquired at a frame interval of 50 ms. **(d)** Raw temporal photon-counting signal recorded by the SPAD, with fluctuations that reflect the droplet growth and dripping cycles. **(e)** Quantitative analysis of the droplet area over time for the sequence in (c), revealing a progressive increase before each dripping event. Scale bar, 5 mm.

## 3. Discussion

In summary, we have demonstrated $HS^3N$, a self-supervised, physics-constrained reconstruction framework designed to mitigate the severe noise of NIR SPI under photon-starved conditions. By integrating deep image prior with a joint regularization strategy which combines $\ell_1$-norm sparsity and Hessian norm, we have successfully reduced the severe Poisson noise that typically degrades ultra-low-light NIR images. Our results indicate that $HS^3N$ outperforms traditional methods, achieving high-fidelity reconstruction at an extreme photon level of ~0.01 photons/pixel at 852 nm. Furthermore, $HS^3N$ maintains reliable reconstruction at a low sampling ratio, enabling the real-time dynamic monitoring of droplet formation at a frame rate of ~20 Hz under sub-photon illumination conditions using a cost-effective, robust and scalable configuration. These findings demonstrate that the proposed method effectively suppresses noise under extremely low illumination, offering a high-fidelity, low-cost, and scalable alternative for NIR sensing.

Despite its strong performance, the effectiveness of the joint regularization strategy is largely dependent on the sparsity of the target scene. Its performance may degrade when dealing with dense, non-sparse, or low-contrast complex scenes. The incorporation of more sophisticated adaptive priors or attention mechanisms could enhance the framework's capacity to resolve non-sparse, complex textures. Looking beyond the NIR band, the $HS^3N$ reconstruction architecture is inherently spectrum-independent and can be readily extended to other challenging regimes, offering a robust, cost-effective, and scalable strategy for visible, short-wave infrared, mid-infrared, or terahertz imaging, where array detectors are either prohibitively expensive or technologically immature. We anticipate that this physics-informed computational approach will pave the way for high-sensitivity, low-cost imaging solutions in fields as diverse as intelligent robotic perception, non-line-of-sight tracking, deep-tissue bio-imaging, and long-range environmental monitoring under extreme conditions.

## Acknowledgments

This work was supported by the National Natural Science Foundation of China (62503032, U23A20481, 62275010, 62573029). We are grateful to the Atomic-Scale In Situ Fabrication Platform of the Analysis & Testing Center of Beihang University for the facilities, and the scientific and technical assistance.

## Conflict of interest

The authors declare that they have no conflict of interest.

## Data Availability Statement

The data that support the findings of this study are available from the corresponding author upon reasonable request.